\documentclass [12pt]{article}
\usepackage {amssymb}
\usepackage {amsmath}
\usepackage{graphicx}
\usepackage {longtable}

\begin{document}
\begin{center}
\textbf{Outlook for development of a scintielectron detector with improved
energy resolution}
\end{center}

\begin{center}
\textbf{\textit{S.V. Naydenov}}\footnote{ E-mail:
naydenov@isc.kharkov.com}, \textbf{\textit{V.D.
Ryzhikov}}\footnote{ E-mail: ryzhikov@stcri.kharkov.ua}
\end{center}

\begin{center}
Institute for Single Crystals, National Academy of Sciences of Ukraine,
\end{center}

\begin{center}
60 Lenin Ave., 61001 Kharkov, Ukraine
\end{center}

\begin{center}
\textbf{Abstract}
\end{center}

The development prospects have been considered of a
scintillator-photo\-diode type detector with improved energy
resolution attaining several per cent ($R = 1-2\%$). The main
contributions to the scintielectron detector energy resolution
have been analyzed theoretically and their theoretical and
physical limits determined. Experimental data have been presented
on properties of scintillators of promise confirming the
possibilities to minimize each of the resolution components. New
ways are proposed to optimize the detector statistical
contribution and the scintillator intrinsic resolution. A special
role of the latter is outlined as the critical factor for the
spectrometric possibilities (threshold) of
scintillation-photodiode type detector with improved energy
resolution at energy values $E_{\gamma}$ from \textit{662 keV} to
\textit{10 MeV}.

\bigskip


\bigskip

Solid state detectors of scintillator-PMT (S-PMT) and
scintillator-photo\-diode (S-PD) types are used widely along with
semiconductor ones (SCD). In the latter type, the ionizing
radiation is converted immediately into charge carriers while in
scintielectron S-PMT and S-PD ones, a two-stage conversion takes
place: first into optical photons (in the scintillator) and then
into charge carriers (in the photoreceiver). The two-stage
conversion causes energy losses and redistribution (dissipation),
thus, in the case of small-volume detectors, the SCD sensitivity
and energy resolution are one decimal order higher. An S-PD
detector comprising a traditional CsJ(TI) scintillator of about
\textit{1sm}$^{3}$ volume and a silicon or HgI2 photodiode has the
energy resolution $5$ to \textit{6\%}  at room temperature for
\textit{662 keV} line [1, 2].

While the energy conversion in scintillators has been considered
comprehensively enough and several models have been proposed to
date describing that process satisfactorily [3-6], there are no
works considering possible correlation between integral
characteristics of that process (quantum efficiency, conversion
one, self-absorption, etc.) and energy resolution of the
scintillator itself and of the scintillator-containing detection
system as a whole; the same is true for theoretical consideration
of the limiting energy resolution as a function of the
scintillator parameters. This work is dedicated to theoretical
consideration of that problem.

Let the energy resolution of a scintillator-photodiode combination
be considered. Under Gaussian approximation of the electron signal
output shape, the resolution $R$ is expressed as

\begin{equation}
\label{eq1} R = \frac{{FWHM}}{{MAX}} = G\frac{\Delta
E}{\overline{E}} \, ,
\end{equation}

\noindent where $G = 2\sqrt {2\ln 2} \approx 2.36$; $\overline{E}$
-- the average energy; $DE \equiv (\Delta E)^{2} =
\overline{E^{2}} - {\overline{E}}^{2}$ -- the spectral line
dispersion. The dependence on the ionizing radiation energy $E$ is
defined by the conversion efficiency of the system, $\eta _{p}$,
in the peak corresponding to the photo-absorption in the
scintillator:

\begin{equation}
\label{eq2} E = \eta _{P} E_{\gamma}  ;\quad \overline{E} =
\overline {\eta }_{P}  E_{\gamma } ;\quad \Delta E = \Delta
E(E_{\gamma} ;\overline {\eta }_{P}  ,D\eta _{P} ;\ldots) \, .
\end{equation}

Strictly speaking, the physical conversion efficiencies of the
system, the total one $\eta _{PD}$ (the energy yield of electrons
per $1 MeV$ of ionizing radiation) and the scintillation one $\eta
_{sc}$ (the absolute light yield per $1 MeV$), and thus the peak
one rip $\eta _{p}$ (the fraction of the total efficiency
corresponding to the photo-absorption peak of the spectral couple
S-PD), are fluctuating being depended not only on the ionizing
radiation energy but also on a multitude of physical and geometric
parameters denoted in (\ref{eq2}) as $(\ldots )$. That is why the
broadening $\Delta E$ (and thus the detector resolution) depends
non-monotonously on its conversion efficiency $R=R(\eta )$,
measured under the current regime. Nevertheless, it follows from
experiments that correlations associated with the light yield
improvement remain conserved and result mainly in an improved
resolution $R,$ in particular, in high-uniform materials
containing no activator.

Mathematically, the non-monotonous dependence $\Delta E = \Delta
E(\eta )$ is due to that the main parameters of Gaussian
distribution, the mean value and dispersion, are explicitly
independent from each other. The spectral line broadening is due
not only to statistical contributions reflecting the relations of
the detector resolution $R$ with its conversion efficiency
\textit{$\eta $} but also to non-statistical fluctuations
including those of geometric, dynamic, space-nonuniform competing
quantities non-proportional to $E_{\gamma }$ which define the
detector conversion yield. Under account for the above
complication, the most effective way to determine the detector
resolution is to discriminate the most important noise channels so
that the corresponding fluctuations could be considered as
independent ones and then to estimate each of those. The
independent fluctuations are added vectorially and result in the
following expression for the resolution (a $\alpha $ is the noise
channel number)

\begin{equation}
\label{eq3}
R^{2} = {\sum\limits_{\alpha}  {R_{\alpha} ^{2}} } {\rm .}
\end{equation}

It is quite natural to define the resolution of a scintielectron
detector (as well as in [8])

\begin{equation}
\label{eq4} {R}^{2}= R_{sc}^{2}+ R_{st}^{2} + R_{pd}^{2} \,,
\end{equation}

\noindent where $R_{sc}$ is the scintillator intrinsic resolution;
$R_{st},$ statistical fluctuations of the energy carriers (photons
and electrons); $R_{pd}, $the noises of photodiode and electronic
devices. Note that each of each contribution in (\ref{eq4}) may in
turn include partial components answering to specific broadening
mechanisms. Since $R_{\alpha}  < R$, it is necessary to provide a
noise level of $1$ to $2\% $  for each component to attain an
improved $R$ value of the same order.

First let conditions be determined allowing to neglect the
photodiode noise $R_{pd}$. This contribution depends on the total
number of noising electrons, $N_{noise}$ related to the useful
electron signal, $N$:

\begin{equation}
\label{eq5} R_{noise} \sim {\frac{{N_{noise}} }{{N}}} =
{\frac{{(\delta E)_{PD}} }{{\eta E_{\gamma} } }} \, ,
\end{equation}

\noindent where $(\delta E)_{PD} $ is the corresponding energy
spread; $\eta $ is meant to be the average value unless otherwise
noted. It will be shown in what follows that a rather high
efficiency of $\eta  \quad \sim  10\%$  is necessary to compensate
the detector statistical noise. From Eq. (\ref{eq5}), we obtain
that, the average energy being $E_{\gamma} = 0.6$ to $1 MeV$ and
the useful signal $N > 30000$ electrons, the resolution $R_{pd}<
1-2\%$  is believed to be small if $(\delta E)_{PD}$ less then
from $l$ to \textit{2keV} or $N_{noise}\sim (2\div 3) 10^{2}$
(electrons). Such a low noise level is already attained in modern
semiconductor photodiodes. For example, the silicon photodiode
used as a SCD has a resolution of \textit{5.5 keV} for \textit{122
keV} line of $^{57}$Co and records reliably \textit{5.5 keV} from
$^{55}$Fe [9]. Thus, its resolution (expressed in relative units)
for $^{137}$Cs line must be about $R_{pd}\sim G(5.5 keV/122 keV)
\sim  0.8\% $, i.e. less than \textit{1\%}. In the high-energy
spectrometry the photodiode noise is even lower, since $N \sim
E_{\gamma }$, in Eq. (\ref{eq5}).

The statistical component $R_{st}$, includes independent
contributions from quantum fluctuations of the number (or energy)
of scintillation photons $R_{st, sc}$ and the photodiode
photoelectrons $R_{st, ph}$ thus,

\begin{equation}
\label{eq6} R_{ST}^{2} = R_{st,ph}^{2} + R_{st,el}^{2} \, .
\end{equation}

In contrast to the cascade theory of statistic fluctuations in a
scintillator-photomultiplier block [10], the statistical
fluctuations of photons cannot be neglected in this case. In
high-resolution scintielectron detectors, a nearly ideal energy
transfer from the scintillation quanta to the photodiode
electrons, therefore, their statistical contributions are of the
same order of magnitude. Note that the $R_{st, ph}$ is often
neglected also in S-PD detectors, but only at resolutions not less
than $4$ to $5\% $.

Let a formula for $R_{st}$ be derived in a system with high
conversion efficiency. Let us consider the conversion of a y
quantum energy into a scintillation photon and then into an
electronic signal as independent events in a Bernoulli check
sequence. The statistical fluctuations of such a check series are
described by binomial distribution [11] with the average value $x
= N p$ and dispersion $Dx = N p (l - p)$ where $N$ is the check
number (the number of converted particles at $100\% $ efficiency);
$p$, the event probability $\eta _{sc}$ and $\eta $,
respectively). Thus,

\begin{equation}
\label{eq7}
R_{sc,\,\alpha}  = G\sqrt {{\frac{{1 - \eta _{\alpha} } }{{\eta _{\alpha}
N_{\alpha} } }}} ;\quad N_{\alpha}  = {\frac{{E_{\gamma} } }{{\varepsilon
_{\alpha} } }}
\end{equation}

\noindent
where $\eta _{\alpha} = (\eta _{sc}, \eta )$;
$\varepsilon _{sc}$ and $\varepsilon _{el}$ are mean energy of the
scintillation quantum and that of the electron-hole pair formation
in the semiconductor, respectively. The total statistical
contribution is

\begin{equation}
\label{eq8} R_{ST} = {\frac{{G}}{{\sqrt {\eta (E_{\gamma}  /
\varepsilon _{el} )}} }}\sqrt{1 + K_{Y} (1 - \eta _{sc}
{\frac{{\varepsilon _{sc} + \varepsilon _{el} }}{{\varepsilon
_{sc}} }})} \, ,
\end{equation}

\noindent where an auxiliary $K_{Y} = (\eta /\eta _{sc})
(\varepsilon _{sc}/\varepsilon _{el})$ is introduced (the
effective collection coefficient). As a rule, $\varepsilon _{sc} =
{{hc} \mathord{\left/ {\vphantom {{hc} {\lambda} }} \right.
\kern-\nulldelimiterspace} {\lambda} } = {{1239} \mathord{\left/
{\vphantom {{1239} {\lambda (nm) \approx 2\div 3\,eV}}} \right.
\kern-\nulldelimiterspace} {\lambda (nm) \approx 2\div 3\,eV}}$;
$\varepsilon _{el} (Si) = 3.6\,eV$. For detectors with low
conversion yield, in the limiting case $\eta \ll \eta _{sc} \ll
1$, the usual Poisson electron noise is obtained from Eq.
(\ref{eq8}):

\begin{equation}
\label{eq9}
R_{ST} \approx R_{st,el}^{poisson} = {\frac{{G}}{{\sqrt {\eta (E_{\gamma}  /
\varepsilon _{el} )}} }}{\rm .}
\end{equation}

At $\eta _{sc} > 20 \% $  and $\eta  > 10\% $ , the difference
between the binomial distribution and the Poisson one become
substantial, the fluctuations of the former drop more sharply than
those of the latter. This evidences that it is reasonable to
enhance $\eta $ and $\eta _{sc}$ for maximum attenuation of the
threshold statistical fluctuations. At the ideal energy
conversion, $\eta =$ $=\eta _{sc}=1$ , the theoretical limit
$R_{st} (\eta = 1)=0$ is attained. An important peculiarity of the
statistical resolution component following from (\ref{eq8})
consists in its monotonous dependence on the conversion
efficiencies. Other contributions to the total resolutions have
been noted above to have not that property.

It is reasonable to transform (\ref{eq8}) so that discriminate
$\eta _{sc}$

\begin{equation}
\label{eq10} R_{st} = {\frac{{G}}{{\sqrt {\eta _{sc} \left(
{E_{\gamma}  / \varepsilon _{el}}  \right)}} }}\sqrt {K_{Y}^{ - 1}
+ \left( {1 - \eta _{sc} {\frac{{\varepsilon _{sc} + \varepsilon
_{el}} }{{\varepsilon _{sc}} }}} \right)} \,.
\end{equation}

Solving this equation with respect to $\eta _{sc}$,

\begin{equation}
\label{eq11} \eta _{sc} = {\frac{{6.91}}{{E_{\gamma}(keV)\lambda
(nm)}}}{\frac{{1 + K_{Y}^{ - 1}} }{{{\left[ {R_{ST}^{2} +
{\frac{{6.91}}{{E_{\gamma} (keV)\lambda (nm)}}}{\frac{{\varepsilon
_{sc} + \varepsilon _{el} }}{{\varepsilon _{el}} }}} \right]}}}}
\, ,
\end{equation}

\noindent we obtain an expression defining the scintillation
efficiency necessary to attain a pre-specified level of
statistical noise in detector having the efficiency $K_{Y}$
defining the matching between the scintillator and the photodiode
($0 < K_{Y} < l$).

When the spectral and optical match is ideal, $K_{Y}$ depends
mainly on the light collection coefficient $K_{c}$, so
$(\varepsilon _{el} / \varepsilon _{sc} ) K_{Y} \sim  K_{c}$. At a
good light collection ($K_{c}$ to $0.8$), the statistical
resolution of a "red" scintillator ($\lambda =640 nm $) on
moderate energy ($E_\gamma = 662 keV $) can be believed to be
small under condition following from equation (\ref{eq11})

\begin{equation}
\label{eq12} R_{ST}(E_\gamma  < 1MeV) < 1\%  \Leftrightarrow \quad
\eta _{sc} \ge  20\div 30\% \,.
\end{equation}

This level is $\eta > 10-15\% $  for the total conversion
efficiency. In high energy range ($E_{\gamma} > 10 MeV$), the
statistical resolution drops sharply in a spontaneous way, since
$R_{st} \propto {{1} \mathord{\left/ {\vphantom {{1} {\sqrt
{E_{\gamma} } }}} \right. \kern-\nulldelimiterspace} {\sqrt
{E_{\gamma} } } }$ . The statistical noise on the \textit{662~keV}
line is $2$ to \textit{4\%}  in best scintillator assemblies [12].
At $E_{\gamma} \sim  10 MeV$, that contribution will be reduced 3
to 4 times to a value less than \textit{1\%}  .

To conclude, it is seen that the statistical contribution can be
minimized essentially in a detector having a high scintillation
efficiency and a good-matched S-PD combination. The first
condition is of high necessity to optimize another contribution
being in our opinion the most important one, namely, the
scintillator intrinsic energy resolution. As a rule,
non-statistical (non-Gaussian) fluctuations are included therein.
The intrinsic resolution $R_{sc}$ is a natural improvement limit
for the total resolution of a detector, since it includes as a
rule contributions remaining substantial (or independent of
$E_{\gamma}  )$ at any energies of the ionizing radiation,
including high ones. The $R_{SC}$ would be a decisive part in
detectors with improved spectrometric resolution $R < 1$ to
\textit{2\%}).

The scintillator intrinsic resolution was noted to be of
importance in the studies of alkali halide scintillators ([3-6]
and other works) where it comes as a rule to $4$ or $5\% $ . In
the case of heavy oxides (BGO and CWO), the intrinsic resolution
of rather small samples ($V\sim  1$ to ${10\, sm}^{3} $) is
negligible due to the proportionality of their light yield [13].
The ZnSe(Te) scintillator intrinsic resolution is $R_{sc}=3.26\% $
at the S-PD couple resolution $R = 5.37\% $ on the $^{137}$Cs line
[14]. The RbGd2Br7:Ce scintillator with proportional light yield
seems to exhibit a rather good intrinsic resolution [15]: the
total resolution on the same \textit{662 keV} line was $R = 4.1\%
$ , while the PMT statistical noise being the main contribution in
that case was $R_{sc} = 3.5\% $. There are no experimental data on
the intrinsic resolution of new scintillators with high values of
atomic numbers $Z$ and $\eta _{sc}$.

The intrinsic resolution consists of several components. Some of
those depend on $E_{\gamma}$, dropping monotonously as a rule.
There are, however, components fully or almost independent of the
ionizing radiation energy if the latter is absorbed completely in
the crystal (the leak resolution or boundary effects being
neglected). Among those threshold contributions, it is just the
scintillator substance resolution $R_{sub}$ and the light
collection non-uniformity one $R_{lc}$ that are the most
substantial ones, that is, at an accuracy to small corrections,

\begin{equation}
\label{eq12*} R_{sc}^{2} = R_{sub}^{2} + R_{lc}^{2} + o(1 / \eta
_{sc} E_{\gamma}  ) \, .
\end{equation}

The substance resolution $R_{sub}$ depends mainly on the light
yield non-proportionality, $E_{sc}=\eta _{sc}(E_{\gamma})
E_{\gamma}$, $\eta _{sc} \neq  const$. The light collection
contribution, $R_{lc}$ , is defined in first turn by the geometric
and dynamic fluctuations in a scintillator of a specific shape,
the optical parameters in the crystal volume and at its boundary
being fixed. The space non-uniformity of scintillations, $\eta
_{sc} = \eta _{sc} (\vec {r})$ can be of importance, in
particular, in activated compounds.

There are scintillators with very low $R_{sub}$\textit{} the light
yield proportionality is their specific feature. The tungstates
mentioned above belong to those. So, CdWO4 has an $R_{sub}$ about
\textit{0.3\%} (as discriminated from the total resolution) at the
crystal volume $V\sim 200 sm^{3}$ and $R_{sub} \quad \approx 0.03$
to $0.08 \% $ at $V \sim  3$ to $20 sm^{3}$ [16]. ZnSe(Te) with
nonlinearity factor $\eta (5.9 keV)/\eta (662 keV) = 85\% $ and
$\eta (16.6 keV) / \eta (662 keV) = 90\% $ has a rather good
linearity at the physical light yield $L = 28000 ph/MeV $ [10].
Some complex oxides are somewhat worse, e.g., Lu3Al5O12:Ce has the
light yield $L = 13000 ph/MeV$ and $\eta (16.6 keV) / \eta (662
keV) = 76\% $ while for LuA1O3:Ce the linearity worsens $\eta
(16.6 keV)/$ $/\eta (662 keV) = 71\% $ as the light yield drops
($L = 11000 ph/MeV$) [17]. For comparison sake, the non-linearity
of NaJ(TI) is about $80\% $ at a light yield $L = 40000 ph/MeV $.
Nevertheless, those compounds, as well as other modern
scintillators, e.g., those of LSO(Ce) type [18], may turn out to
be of promise to attain a high energy resolution of scintillators,
including both total and intrinsic one, as their scintillation
characteristics will be further improved.

To minimize the substance contribution, a material should be
developed having a high scintillation efficiency, homogeneity and,
most likely, an intrinsic emission. Since the conversion process
in scintillator is a multi-factor process, there is no functional
relation between $\eta $, and $R_{sub}$ . The above examples show,
however, that there is such a correlation for materials of the
same type while it is not confirmed when materials of
substantially different types are compared. For example, CdWO4 has
the light yield 30 \% related to CsJ(TI) but their total
resolutions are comparable [19] due to that the intrinsic
resolution of the former is much lower (2 to 3 decimal orders).

The space uniformity of scintillations should be high enough along
with their proportio\-nality ($R_{sub,nonprop} \ll 1\% $).
Corresponding residual resolution (as defined in [6])

\begin{equation}
\label{eq13}
R_{sub,in} = 2.36{\frac{{\sqrt { < \eta _{sc}^{2} > - < \eta _{sc} > ^{2}}
}}{{ < \eta _{sc} >} }};
\quad
 < ... > = {\frac{{1}}{{V}}}{\int\limits_{V_{sc}}  {d\vec {r}}} ...
\end{equation}

\noindent is small, $R_{sub,in} < 1\% $, under condition that
$(\Delta \eta _{sc}) < 10^{ - 4} \langle\eta _{sc}\rangle $. That
condition will be met in systems where fluctuations are very
small, e.g., in a regular lattice. The known method to compensate
the $\eta _{sc}$ non-uniformity in an activated scintillator by
providing the light collection non-uniformity (the multiplicative
noise of $\langle \eta _{sc}K_{c}\rangle \neq \langle \eta _{sc}
\rangle \langle K_c \rangle $ value) is unsuitable in this case
because it results in a definite drop of $\langle K_{c}\rangle $
and thus of the conversion efficiency that is inadmissible in
high-resolution detectors.

Now let geometric fluctuations of light collection, $R_{lc}$, be
considered. At the uniform scintillation distribution $R_{lc}$ is
due only to the dispersion of light collection coefficient and, in
spite of the non-statistical fluctuation character, has the
following form under Gaussian approximation:

\begin{equation}
\label{eq14} R\,_{\sigma}  = 2.36{\frac{{\sqrt { < K_{c}^{2} > - <
K_{c} > ^{2}}} }{{ < K_{c} >} }} \, .
\end{equation}

The light collection parameters are expressed via the invariant
distribution function $\rho $ for the totality of light beams
being reflected from boundary $\partial \Omega $ [20]. The
averaged $\langle K_{c}\rangle $ is

\begin{equation}
\label{eq15}
 < K_{c} > = 1 - {\int\!\!\!\int\limits_{\Phi _{catch}}  {\rho (\varphi _{1}
,\varphi _{2} )d\varphi _{1} d\varphi _{2}} } ; \quad \vec
{r}(\varphi ) \in \partial \Omega \, ,
\end{equation}

\noindent where the integration is made over the part of the
system phase space $\Phi _{capt}$ corresponding to the captured
light.

The light collection resolution, as well as the coefficient
$K_{c}$ itself, depends on the scintillator geometry and optical
properties (the light reflection, refraction and absorption). When
the light collection is ideal, the reflection at the boundary is
the mirror one and there is no light absorption in the
scintillator, the resolution is minimum, $R_{lc}< 1 \% $. Strictly
speaking, the theoretical limit $R_{lc} = 0$ is attained in this
case, that is confirmed by several authors (see e.g. [16]). The
reason for the ideal character of mirror light collection is
established within the frame of stochastic (geometry-dynamic)
light collection theory [20]. In this case, a dynamic model
presenting a billiard with elastic reflections at the boundaries,
is considered instead of the detector. There are no fluctuations
for mirror light collection, $<K_{c}>^{2}$ = $<K_{c}>^{2}$, due to
the unique (deterministic) character of the light propagation.
Therefore,

\begin{equation}
\label{eq16} R_{lc,mirror} (\kappa = 0) \equiv 0 \,.
\end{equation}

(where $\kappa $ is absorption coefficient). There is a quite
different situation when the absorption takes place. It results
not only in reduced scintillation intensity but also causes a
light collection non-uniformity. The effective light collection
with a high resolution is attained in small-size and/or optically
transparent scintillators. It follows from experiments, numerical
calculations and general considerations that the light collection
resolution is small enough ($R_{lc}\ll 1 \%$) or a regular
geometry scintillator of a volume $V $about $10\, sm^{3}$ (e.g.,
for a cylinder with $H \sim  D \sim 3sm $) and having
mirror-reflecting boundaries (with mirror efficiency $\rho _{m}
\quad \sim  \quad 0.8$ or $0.9$ ) under condition that

\begin{equation}
\label{eq17}
R_{lc} < 1\% \Leftrightarrow \kappa d \le 0.1\;\;\;\;(10d \le l),
\end{equation}

\noindent where $\kappa $ is the optical absorption coefficient;
$l = \kappa ^{-1}$ , the light beam free path in the scintillator;
$d$ -- the crystal characteristic dimension. A good light
collection, $K_{c}\sim 60\% $, is attained under the same
conditions.

To conclude, a convenient determination of $R_{sc}$ from the total
one, $R$, measured at several $E_{\gamma}$ values is described in
what follows. Since $R_{st}^{2} = const / E_{\gamma} $ and the
total $R$ is determined from (\ref{eq4}), we obtain at known

\begin{equation}
\label{eq18} R_{sc} = \sqrt {{\frac{{R_{1}^{2} E_{1\gamma}  -
R_{2}^{2} E_{2\gamma} }}{{E_{1\gamma}  - E_{2\gamma} } }} -
R_{PD}^{2}}  \quad ,
\end{equation}

For small-size samples where $R_{lc} \ll 1\% $, $R_{sc}\sim
R_{sub}$. The scintillator intrinsic resolution is minimum for
detectors spectral line a pronounced Gaussian character of, i.e.
$R^{2} \sim 1/E_{\gamma}$.

For medium energies, $E_{\gamma} < 1 MeV$, the energy resolution
limit depends on the statistical fluctuation level. To attain
$R_{st}$ to $2 \% $, it is necessary to use scintillators with
high conversion efficiency ($\eta _{sc}\sim  20 \%$) and good
matching (both spectral and optical) between S and PD as well as
to optimize the light collection coefficient so that the total
efficiency $\eta $ would be not less than $10 \%$.

For the high energy range, $E_{\gamma} > 10 MeV $, it is the
intrinsic energy resolution of the scintillator that is decisive.
It consists of several components which are in general competing
with each other. First of all, scintillators are to be sought
having a vanishing resolution of the material independent of the
crystal dimensions and being defined by the light yield
non-proportionality due to the emission mechanism itself.
Moreover, the scintillator size (volume) is to be restricted to
minimize non-uniformity in the light collection and in the optical
absorption factor. This may result in reduced intrinsic resolution
caused by the light yield non-proportionality associated with the
ionizing radiation incomplete absorption (leaks) and by increased
contribution from micro\-scale inhomogenates (in activated
scintillators). The leak resolution can be avoided if radiation
weakening of y emission in the scintillator is complete and the
secondary electron path length therein is small enough, thus,
under conditions ($2$ to $3$) $l_{r} < d$ , where $l_{r}\sim
Z^{p}$ ($p=2\div 5$) is the scintillator radiation length, and $10
l_{e} < d $, where $l_{e} \sim 0.45 E (MeV)/\rho _{sub}(g/sm^{3})$
is the electron free path in the scintillator. A conclusion of
importance follows therefrom that the intrinsic $R_{sc}$ will be
reduced considerably in scintillators with high $Z$ and $\eta $
values.

It is of interest to compare experimental data on energy
resolution at $\alpha $, $\beta $-particles and $\gamma
$-radiation recording using a modern scintillator with high
conversion efficiency (ZnSe-Te). The energy conversion for $\beta
$ and $\gamma $-radiation takes place in essentially the same
range (for $\alpha $ in somewhat higher one). The conversion
efficiency, however, is close to \textit{100\%}  for $\alpha $ and
$\beta $ particles while it is less than \textit{1\%} for $\gamma
$-radiation, the scintillator thickness being about $1 mm$ in all
cases. That is, the complete energy absorption, light conversion
optimization and low statistical noise are simulated correctly for
$\alpha $ and $\beta $ particles in a high conversion efficiency
scintillator. In contrast, as y radiation is detected, the
photoreceiver and electronic noises are of substantial importance
due to low conversion efficiency. It is seen (Fig.l) that the
experimental result for $\alpha $ and $\beta $ detection using a
S-PD couple coincides essentially with SCD data being several per
cent ($4$ to $6\%$).

ZnSe has a low transparency, thus, it is unreasonable to use the
large crystals for high-efficiency absorption of y radiation. The
energy resolution of thin scintillators on \textit{662 keV} line
(Fig.l), however, confirms, under account for the predominant part
of PD and electronic noises, the attainability of a high energy
resolution using an S-PD combination using a homogeneous
scintilla-tor transparent against intrinsic emission and showing a
high conversion efficiency as well as a low intrinsic resolution.
To compare (see Fig. 2), the energy resolution for standard
CsJ(Tl) is less than \textit{6\%}  on \textit{662 keV} line and
less than \textit{5\%}  on \textit{1.33 MeV} one at room
temperature.

In high energy range, $E_{\gamma} > 10 MeV$, all requirements
concerning the medium energy range remain valid. Under account for
the obligatory high detector efficiency, i.e., the high-energy
radiation absorption coefficient of several tens per cent, it is
just high intrinsic resolution, homogeneity and light collection
optimization that become the decisive factors. The resolution
reduction down to \textit{1\%}  and less is illustrated (Fig.~3)
for lead tungstate detectors (\textit{GeV} energy range). In
practice, this means that non-activated scintillators of high $Z$
and small radiation length should be used having the emission in
"red" spectral range providing a better matching with the
photodiode. Note that in the high energy range the same absorption
efficiency is attained at the BGO or PWO scintillator volume 8
times smaller than that of CsJ or NaJ. The improvement of the
conversion efficiency for high atomic number scintillators is here
the main problem.

The theoretical analysis allows to conclude that it is quite
possible physically to attain sufficiently low values of $R_{PD}$,
$R_{ST}$ and $R_{SC}$ components of S-PD detector energy
resolution (except for low-energy range where the statistical
contribution predominates and increases sharply as the radiation
energy drops). Thus, it is possible to attain $R = 1\%$  to $2\%$
. The necessary conditions for such a small spectral line
broadening (as compared to existing scintielectron detectors
having $R = 4$ to $5\%$) are a considerably improved conversion
efficiency (by a factor 1.5 to 2) and scintillation one (by a
factor of 2 to 3, i.e., up to $\eta _{sc}\sim 25-30\%$). It is
necessary also to search for optically transparent homogeneous
scintillators close to intrinsic emission exhibiting a low
non-proportionality of light yield (small intrinsic resolution)
and high atomic number.

The modern advances in production of high $Z$ crystals exceeding CsJ(TI) in
light yield [21,22] evidence the practical solvability of that task.

\bigskip

\begin{center}
\textbf{\textit{References}}
\end{center}

[1] Gramsch, K. Lynn, M. Weber et al., \textit{Nucl.
Instr.Meth.,}\textbf{ A 311,} 529 (1992).

[2] J.M. Markakis, \textit{Nucl. Instr.Meth.,}\textbf{ A 263,} 499
(1988).

[3] G.G. Kelly, P.K. Bell, R.C. Davis, N.H. Lazar, \textit{IEEE
Trans. Nucl. Sci.,} NS-3, 57 (1956).

[4] P. Iredale, \textit{Nucl. Instr.Meth.,}\textbf{ 11,} 340
(1961).

[5] C.D. Zerby, A. Meyer, R.B. Murray, \textit{Nucl. Instr.and
Meth.,}\textbf{ 12,} 115 (1961).

[6] R. Hill, A.J.L. Collinson, \textit{Proc. Phys. Soc.,} 85, 1067
(1965).

[7] W. Zimermann, \textit{Rev. Scient. lustrum.,}\textbf{ 32,}
1063 (1961).

[8] J.M. Markakis, \textit{IEEE Trans. Nucl. Sci.,}\textbf{
NS-35,} 356 (1988).

[9] H. Grossman, \textit{Nucl. Instr. and Meth.,}\textbf{ A 295,}
400 (1990).

[10] E. Bretenberger, \textit{Progr.Nucl.Phys.,}\textbf{ 10,} 16
(1955).

[11] W. Feller, \textit{An Introduction to Probability Theory and
its Applications}, vol. l. New York (1970).

[12] L.V. Atroshchenko, L.P. Gal'chinetskii, V.D. Ryzhikov et al.,
\textit{Scintillator Crystals and Detectors of Ionizing Radiations
on Their Base}, Naukova Dumka, Kyiv (1998) [in Russian].

[13] P. Dorenbos, M. Marsman, C.W.E. van Eijk, in: Inorganic
Scintillators and their Applications, ed. by P.~Dorenbos, Proc.
Intern. Conf., Delft, the Netherlands (1995).

[14] M. Balcerzyk, W. Klamra, M. Moszynski et al., Intern. Conf.
SCINT-99, Book of Abstracts, August 16-20, 1999, Moscow, p.125

[15] O. Guillot-Noel, J.C. van't Spijker, P. Dorenbos et al.,
Intern. Conf. SCINT-99, Book of Abstracts, August 16-20, 1999,
Moscow, p.108.

[16] Yu.A. Tsirlin, M.E. Globus, E.P. Sysoeva,
\textit{Optimization of Detecting Gamma-Irradiation by
Scintillation Crystals}, Energoatomizdat, Moscow (1991) [in
Russian].

[17] M. Balcerzyk, W. Klamra, M. Kapusta, SCINT-99, Book of
Abstracts, August 16-20, 1999, Moscow, p.61; W.W. Moses, M.J.
Weber, S.E. Derenzo et al., Intern. Conf. SCINT-97, Book of
Abstracts, September 22-25, 1997, Shangai, China, p. 358.

[18] M. Moszynski, M. Kapusta, D. Wolski et al.,\textit{ IEEE
Trans. Nucl. Sci.,}\textbf{ 45,} 472 (1998).

[19] S. Burachas, Yu. Zdesenko, V. Ryzhikov et al., \textit{Nucl.
Instr. and Meth.,}\textbf{ A 369,} 164 (1992).

[20] S.V. Naydenov, V.V.Yanovsky, \textit{Functional
Materials},\textbf{ 7,} \# 4, 743 (2000); \textbf{8}, \# ~1, 27;
\# 2, 226\textbf{} (2001).

[21] W.W. Moses, M.J. Weber, S.E. Derenzo, P. Perry, P. Berdahl,
L. Schwarz, U. Sasum, L.A. Boatner, Intern. Conf. SCINT-97,
September 22-25, 1997, Shangai, China, p. 358; J.C. van't Spijker,
P. Dorenbos, C.P. Allier, C.W.E.van't Eijk, A.E.Ettema, Ibidem,
Shangai, China, p. 311.

[22] E.V.D. van Loef, P.Dorenbos, C.W.E. van Eijk, \textit{Appl.
Phys. Lett.}, \textbf{77}, \# 10, 1467 (2000); \textit{Appl. Phys.
Lett}., \textbf{79}, \# 10, 1573 (2001).

\newpage

\begin{figure}
\centering
\includegraphics[width=4.44in,height=3.61in]{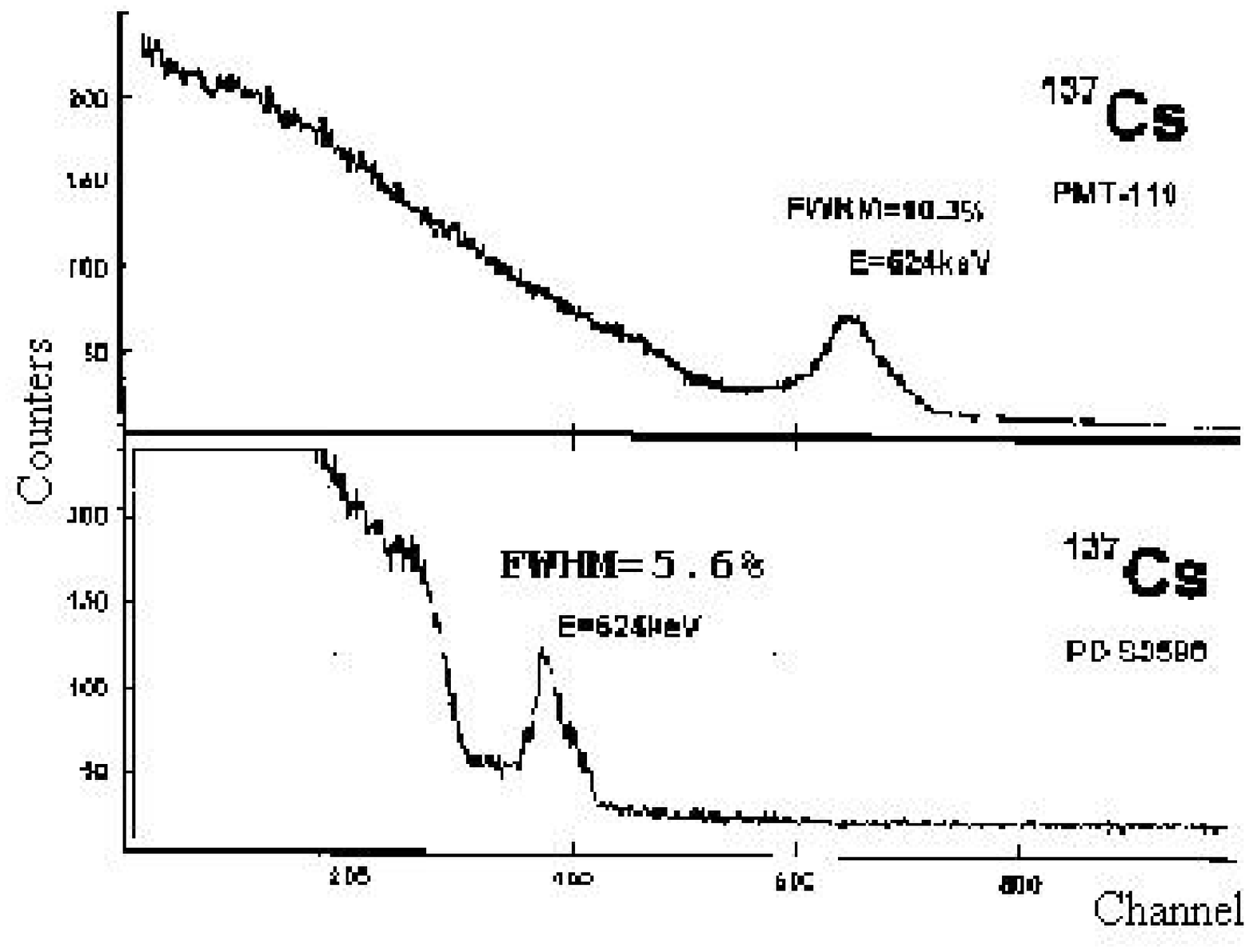}
\end{figure}
\textbf{Fig. l}. Spectrum of $^{137}$Cs conversion electrons
recorded by an S-PMT couple (a) and an S-PD one (b). Scintillator
ZnSe(Te), $1 mm$ thick.

\newpage

\begin{figure}
\centering
\includegraphics[width=5.45in,height=5.52in]{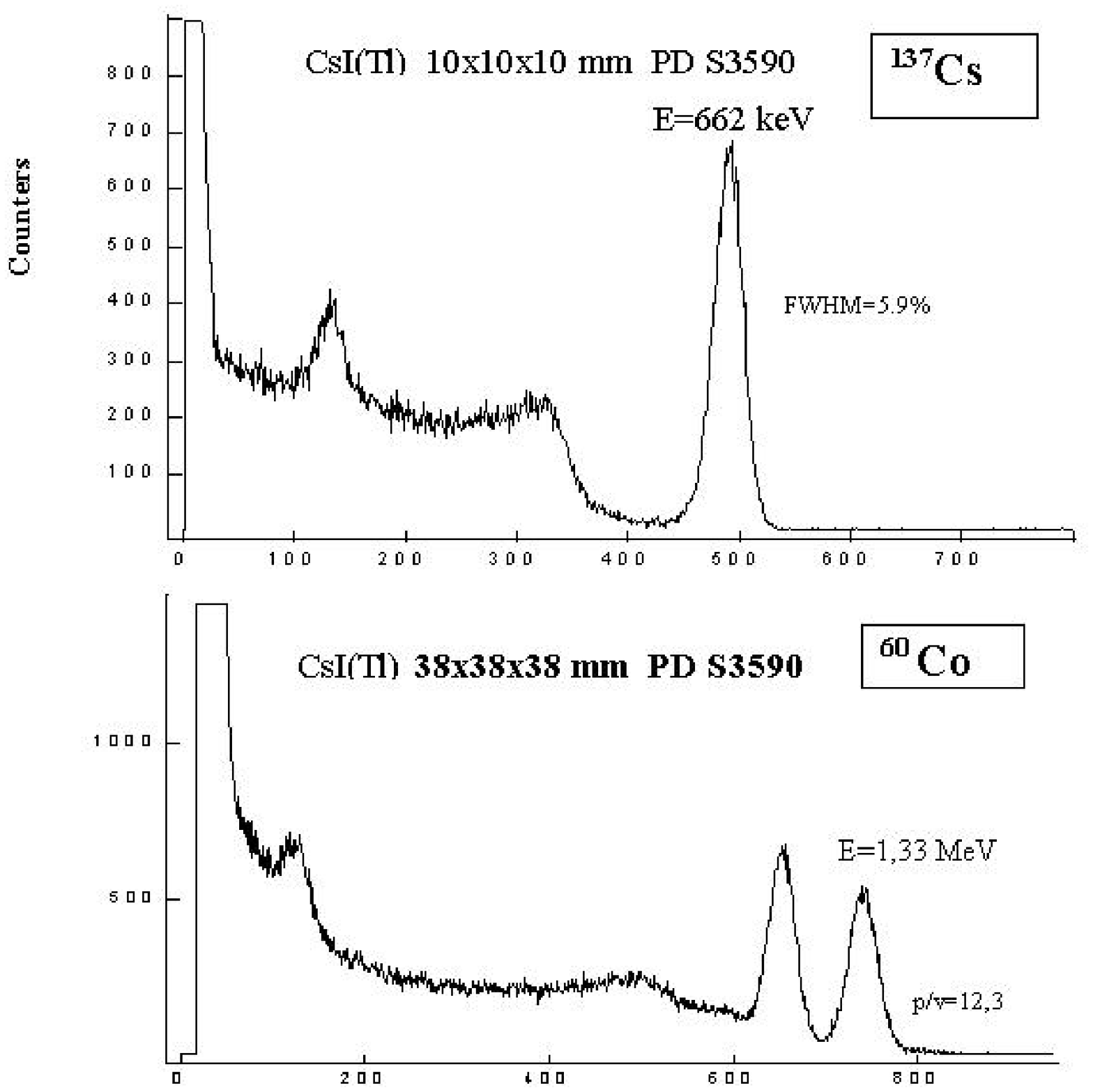}
\end{figure}

\textbf{Fig. 2}. $^{137}$Cs (a) and $^{60}$Co (b) radiation
spectra detected S-PD (CsJ(TI) and PD S3590); scintillator size:
a) $10{\rm x}10{\rm x}10 mm^{3}$, b) $38{\rm x}38{\rm x}38
mm^{3}$.

\newpage

\begin{figure}
\centering
\includegraphics[width=4.05in,height=3.95in]{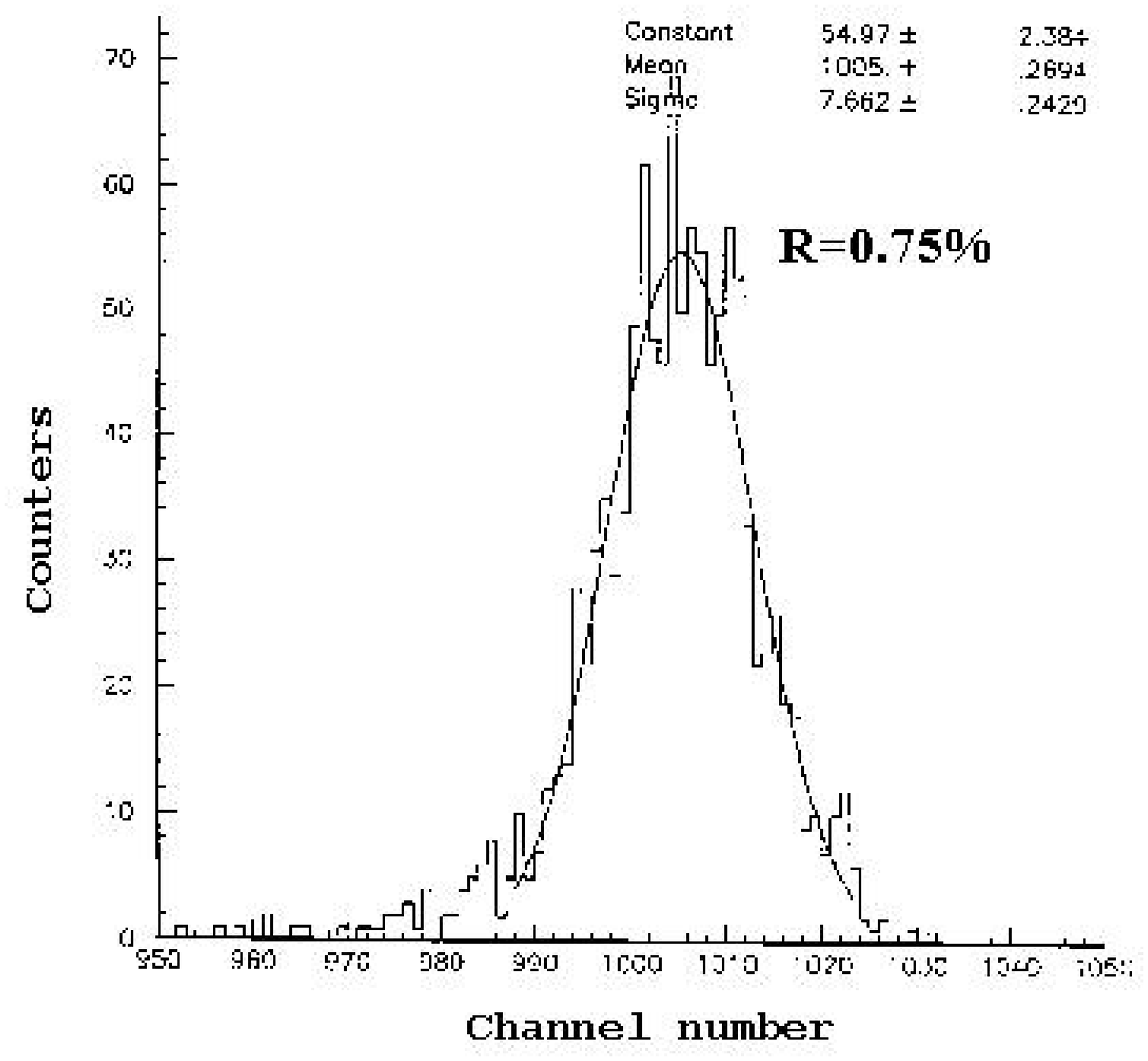}
\end{figure}

\textbf{Fig. 3}. $100 GeV$ radiation spectra detected by S-PD
assembly with a PbWO4 scintillator and PD S3590 potodiode.
Resolution less than $0.75 \%$ in both cases.

\end{document}